\documentclass[manuscript,nonacm,screen]{acmart}
\usepackage{enumitem}

\AtBeginDocument{%
  \providecommand\BibTeX{{%
    \normalfont B\kern-0.5em{\scshape i\kern-0.25em b}\kern-0.8em\TeX}}}
\setcopyright{acmlicensed}
\copyrightyear{2018} 
\acmYear{2018}
\acmDOI{XXXXXXX.XXXXXXX}

\begin{document}

\title{Quality assessment of brain structural MR images: Comparing generalization of deep learning versus hand-crafted feature-based machine learning methods to new sites
}

\author{Prabhjot Kaur}
\authornotemark[1]
\email{prabhjot.kaur@childrens.harvard.edu}
\affiliation{%
  \institution{Hawkes Institute and Department of Computer Science, University College London}
  \city{London}
  \country{United Kingdom}
}



\author{John S. Thornton}
\affiliation{%
 \institution{Neuroradiological Academic Unit, UCL Queen Square Institute of Neurology, University College London}
 \city{London}
 \country{United Kingdom}}
\affiliation{%
  \institution{Queen Square Centre for Neuromuscular Diseases, Department of Neuromuscular Diseases, UCL Queen Square Institute of Neurology, London}
  \city{London}
  \country{United Kingdom}
}
\author{Frederik Barkhof}
\affiliation{%
 \institution{Neuroradiological Academic Unit, UCL Queen Square Institute of Neurology, University College London}
 \city{London}
 \country{United Kingdom}}
\affiliation{%
 \institution{Neuroradiological Academic Unit, UCL Queen Square Institute of Neurology, University College London}
 \city{London}
 \country{United Kingdom}}
\affiliation{%
 \institution{Radiology \& Nuclear Medicine, VU University Medical Center}
 \city{Amsterdam}
 \country{Netherlands}}

\author{Tarek A. Yousry }
\affiliation{%
 \institution{Neuroradiological Academic Unit, UCL Queen Square Institute of Neurology, University College London}
 \city{London}
 \country{United Kingdom}}
\affiliation{%
 \institution{Queen Square Centre for Neuromuscular Diseases, Department of Neuromuscular Diseases, UCL Queen Square Institute of Neurology}
 \city{London}
 \country{United Kingdom}}

\author{Sjoerd B. Vos }
\affiliation{%
  \institution{Hawkes Institute and Department of Computer Science, University College London }
  \city{London}
  \country{United Kingdom}
}
\affiliation{%
 \institution{Neuroradiological Academic Unit, UCL Queen Square Institute of Neurology, University College London}
 \city{London}
 \country{United Kingdom}}
\affiliation{%
 \institution{Western Australia National Imaging Facility node, The University of Western Australia }
 \city{Perth}
 \country{Australia}}

 \author{Hui Zhang}
\affiliation{%
  \institution{Hawkes Institute and Department of Computer Science, University College London }
  \city{London}
  \country{United Kingdom}
}




\begin{abstract}
      
\end{abstract}



\maketitle

\section{Introduction}
Today automated image analysis is routinely used for information extraction from brain Magnetic Resonance (MR) images.  However, increasingly its performance has been found to be strongly affected by the quality of input images, for a range of MR modalities (See e.g.~\cite{REUTER2015107,Klauschen2009,Alexander,power2012spurious,yendiki2014spurious,haller2014head}).  These studies confirm that poor quality images, which may stem from faulty hardware or patient compliance issues, if not identified and excluded, may produce erroneous biomarker estimates.

In this context, of particular importance is the case of high-resolution T1-weighted (T1w) structural MRI.  There are two reasons for this. First, structural MRI is widely used in neuroimaging, as it is the modality of choice for quantifying brain morphological changes.  Second, it is also especially prone to motion artifacts.  A number of independent studies have demonstrated consistently that structural MRI data corrupted by motion artifacts can cause automated analysis methods to produce biased assessment, e.g. underestimating gray matter volume and cortical thickness~\cite{REUTER2015107,Alexander}.

To avoid such erroneous outcomes, it is important that structural MRI scans are assessed for their quality, so that only those with sufficient quality are included for automated analysis.  Currently, image quality is typically assessed through visual rating.  However, this approach is time-consuming and subjective, requiring trained raters with sufficient expertise to minimise intra- and inter-rater variabilities.  These requirements make the approach largely infeasible, or costly to implement, for large-scale studies, such as the UK Biobank imaging study~\cite{sudlow2015uk,citenature}.

To provide an alternative to visual rating that is more scalable and less operator-dependent, the community has been working towards automated quality assessment (AQA) powered by machine learning.  The initial efforts focused on identifying effective image-quality metrics (IQM) that can be automatically computed from images~\cite{Woodard2006,aqaobjective}.  These developments were necessary because the machine learning (ML) techniques, before the recent emergence of deep learning (DL), required inputs, known as features, that are significantly more compact than the images themselves.  IQMs have now been successfully utilised as the required input features in a number of AQA tools~\cite{mriqc,aqaSVM,alfaro2018image}; as IQMs are human engineered, we refer to them as hand-crafted.  However, AQA tools that require IQMs have one significant limitation. They are often computationally expensive, because the automated estimation of IQMs can involve substantial pre-processing (e.g., tissue segmentation)~\cite{mriqc}. Additionally, the existing IQMs may not capture all of the discriminative image characteristics.

The more recent AQA tools have adopted DL to mitigate these inherent limitations of the existing IQM-based methods~\cite{AbdominHead,liver,JMRISheeba,warehouse,3DCNNMRART}.  Compared to the conventional ML techniques, a key strength of DL is that there is neither the need to know {\it a priori} a set of hand-crafted features nor the need to pre-compute them from input images.  DL allows us to take a part or the whole of an image as the input and learn the features appropriate for any given task automatically~\cite{goodfellow2016deep}.  Developing DL-based AQA approaches, however, is not without its own challenges.  Like most DL-based methods, the ability to learn its own features comes with a price: these methods generally demand significantly more data to train and the training data must be labeled, i.e., we need to know if an image used in training is good quality or not.  As visual rating must be used to determine the label, the curation of such dataset can be costly, likely more so for DL-based methods than for IQM-based ones.

Given their differing strengths and weaknesses, choosing one method over the other in practice will likely be challenging.  To help inform their relative merit among practitioners, there is a need for studies designed to compare these approaches in real-world settings.  One recent study did exactly that, comparing MRIQC~\cite{mriqc}, a publicly available IQM-based method, against a DL-based method implemented with a 3D convolutional neural network (CNN) architecture~\cite{3DCNNMRART}, finding that the two approaches perform similarly in terms of balanced accuracy.  The authors thus suggested that DL-based approaches should be preferred over IQM-based methods due to the latter's need for computationally expensive IQM estimations.  However, one limitation of this comparison~\cite{3DCNNMRART} is that the data used for performance evaluation came from some of the sites that also contributed the data for training.  This means that its conclusion may not inform arguably the more common scenario, likely the one preferred by the practitioners, in which off-the-shelf AQA tools must be deployed at new sites without the possibility to improve the tools, due to a lack of site-specific training data to improve the tools or expertise.

Our study aims to address this gap by extending the previous comparison~\cite{3DCNNMRART} to incorporate a leave-one-site-out evaluation.  Such an evaluation was previously used to assess MRIQC~\cite{mriqc} to understand its effectiveness across diverse sites.  Here we adopt their approach but for comparing IQM- and DL-based methods.  As in~\cite{3DCNNMRART}, MRIQC~\cite{mriqc} is chosen as the representative of IQM-based methods, because it represents the state-of-the-art in this category and is publicly available.  To represent DL-based methods, we have chosen the more accessible 2D CNN architecture of CNNQC~\cite{JMRISheeba}, as it has significantly less memory requirement compared to 3D CNN architectures used in other AQA works - with 2.5\% of number of parameters of the 3D CNN in~\cite{3DCNNMRART}.  The objective of this work is to determine the relative performance of these two family of methods for unseen images from (i) same sites used for training, and (ii) new sites not used for training.

\section{Materials and Methods} 
This section describes the dataset we identified as the most suitable for the proposed comparison, implementation of the methods, and evaluation strategy that ensures a fair head-to-head comparison between the methods. 




\subsection{Dataset}
The dataset appropriate for this work shall have (i) volumes acquired at multiple sites, (ii) expert quality labels for each volume, (iii) option to access volumes and quality labels with minimal restrictions. We chose ABIDE\footnote{https://fcon\_1000.projects.nitrc.org/indi/abide/}~\cite{abide} which  satisfies above requirements and has also been used to evaluate image quality by both MRIQC and CNNQC methods. 

\subsubsection{Images}
\begin{figure}
    \centering
    \includegraphics[width=0.95\linewidth]{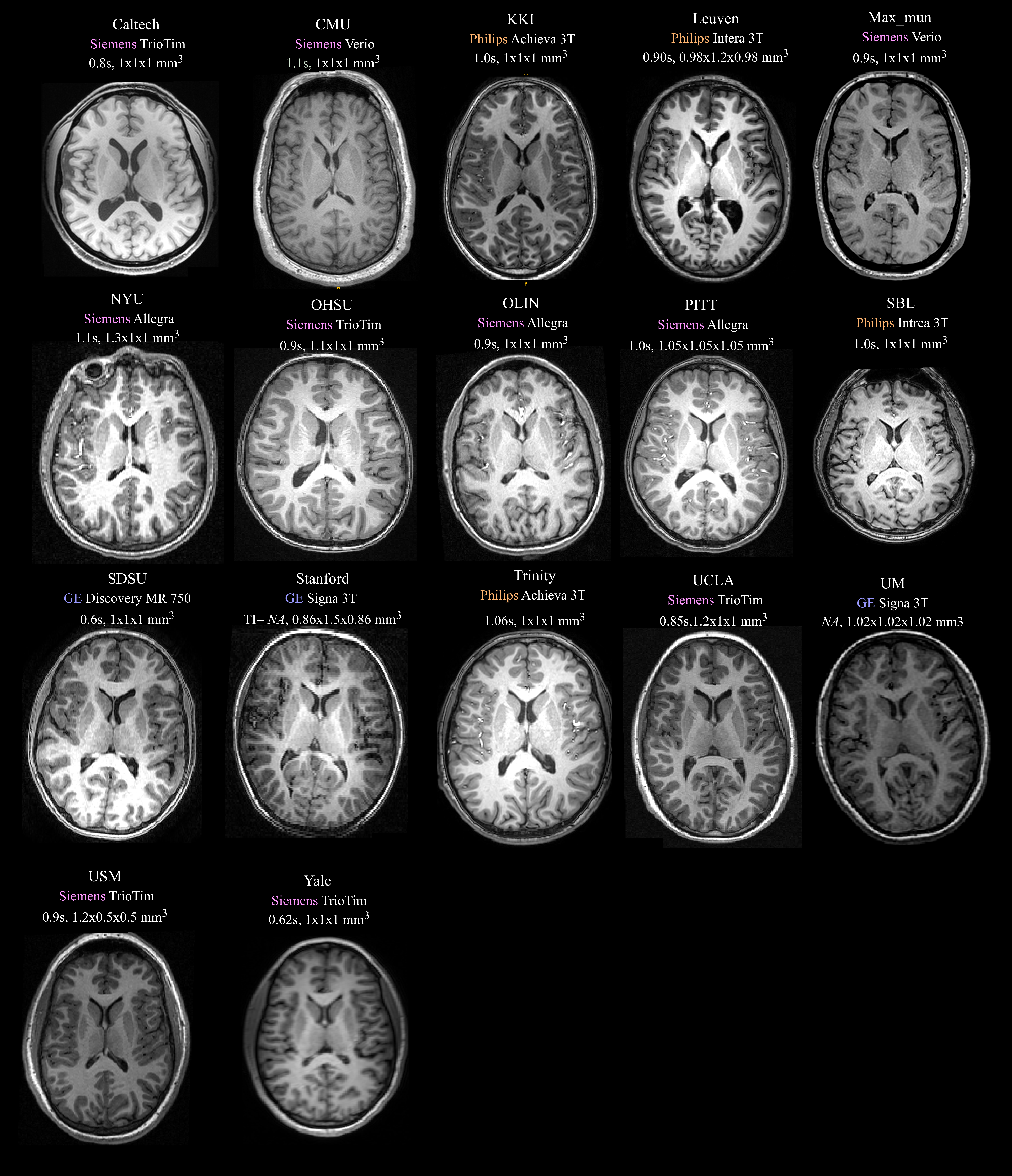}
    \caption{Example T1-weighted MRI images from 17 different sites in the ABIDE dataset, 
    spatial resolution, and MRI scanner vendors used during acquisition.}
    \label{Figure:17sites}
\end{figure}

Following MRIQC~\cite{mriqc}, we used 1102 high-resolution 3D T1w volumes from ABIDE dataset acquired at 17 different sites with 7 different MRI scanner models from 3 different manufacturers. The structural images were acquired with vendor-specific acquisition protocols and acquisition parameter values which differ in each site~\cite{mriqc}. The age range of the subjects differed across sites from 6.47 to 64 years old (median 14.66 years). Hence, this dataset is heterogeneous with images as shown in Fig.~\ref{Figure:17sites} of varying brain anatomy (children to elderly), resolution (voxel sizes = 1.13 $\pm$ 0.27 mm $\times$ 1.0 $\pm$ 0.5 mm $\times$ 0.93 $\pm$ 0.43 mm) and variations in image contrast in T1w images, and thus a good representation of a real-world scenario, establishing it as an appropriate choice for this study. 

\subsubsection{Labels}
There are two sets of publicly available image quality labels for the ABIDE dataset. The first set is included in the phenotypic information of the ABIDE dataset itself~\cite{abide}. The second set, derived from the MRIQC study, involves expert labeling conducted by the MRIQC authors, who provided expert quality labels for the ABIDE dataset based on accuracy of surfaces generated from images, signal to noise ratio, motion etc~\cite{mriqc}. The volumes were labeled as good/accept, borderline/doubtful or poor/exclude. We used labels from~\cite{mriqc} aligning our work with previous MRIQC research for consistency and comparability.  

Following MRIQC, we grouped the images of good and borderline classes into single class `good' to make quality prediction as binary classification problem. 
In MRIQC ratings, each image was assessed by at least one rater and 100 images were assessed by two raters. It was demonstrated in MRIQC~\cite{mriqc} that inter-rater agreement was improved (from $\kappa=0.38$ to $\kappa=0.51$) when borderline class was combined with `good' class. By reducing inter-rater variability, the likelihood of label noise in the data can be minimized, which is known to otherwise effect the performance of supervised classification methods~\cite{goodfellow2016deep}. 
Following this, we labeled the images, those were labeled by more than one rater, as bad if at least one rater has labeled it as bad. 

One volume 
with poor outcome  (missing cerebellum) for pre-processing executed for minimally pre-processed ABIDE-I dataset, and three volumes with ambiguous quality labels were excluded in this study, resulting in a final dataset of 1098 scans for analysis. 
The numbers of good and bad quality images were 756 and 342, respectively. This resulted in class imbalance where good-quality images were twice as numerous as poor-quality ones. Class imbalance needed to be taken into account while designing training and evaluation strategy in different methods to avoid trivial solutions and achieve accurate outcomes.  




\subsection{Methods Under Evaluation}
This section presents the key implementation details of MRIQC and CNNQC~\cite{mriqc,JMRISheeba}, with a focus on aspects that underscore their methodological differences. We also highlight the rationale behind  specific modifications introduced during implementation of methods in this work. 

\subsubsection{MRIQC}
This method consists IQM features, and training support vector machine (SVM) on the IQM features in a supervised manner with area under curve (AUC) as a loss function addressing the class imbalance present in dataset. The source code of MRIQC is publicly available and was used in this work without any modifications, employing default settings to generate quality predictions for input volumes\footnote{https://github.com/nipreps/mriqc}. 



\subsubsection{CNNQC}

We used the same implementation as suggested in ~\cite{JMRISheeba}, with slight modifications in pre-processing, and network architecture. The volumes were not interpolated or conformed to 1mm$^3$ isotropic resolution as done in ~\cite{JMRISheeba} to avoid interpolation effects. We used a single CNN trained on axial slices rather than three separate CNNs for axial, coronal, and sagittal planes because the multi-plane approach yielded only negligible performance gains.

\subsection{Evaluation Strategy}


    
\subsubsection{Evaluation Strategy.} 
We compared MRIQC and CNNQC methods for different levels of generalization by designing test cohort disjoint on : (i) subject level - evaluate on unseen images from sites which were used in training phase, (ii) site level - evaluate on unseen images from the sites which were held out in training phase. This leads to two evaluation strategies: (i) seen site and (ii) unseen site evaluation. 
For fair head-to-head comparison in both evaluations, train+validation and test cohorts were kept the same for both MRIQC and CNNQC methods. 

\subsubsection{Constructing Training and Test cohorts}
\label{sec:dss}
Since the two kinds of evaluation strategies differ in criteria to select images and sites in training paradigm, we design two separate data splitting schemes i.e., categorizing volumes  into train/validation/test cohorts. 

\begin{enumerate}[label=(\roman*)]
    \item \textit{For seen site evaluation: } All images from all sites were used in each of the train, validation, and test cohorts. The ratio of 60\% for training, 20\% for validation, and 20\% for test cohort was maintained while repeating data splitting 5 times.

\item \textit{For unseen site evaluation: } A single site was selected and designated as the test site and all its images were categorized as the test cohort. The image volumes from the remaining 16 sites were randomly shuffled and assigned to the train and validation cohorts, maintaining the same ratio of good vs. poor quality images for each site in the training and validation cohorts. This process was repeated 17 times to evaluate the performance for all 17 different sites.

\item \textit{Data stratification strategies for both seen and unseen site evaluation: }
To ensure unique train/validation cohorts across cross-validation folds, prevent data leakage from training/validation to test sets, and to ensure that the validation cohort accurately represents the training distribution, images were stratified into training, validation, and test cohorts for each of the two evaluation schemes as follows:

\begin{itemize}
    \item Randomly shuffle subjects in each site before constructing training/validation/test cohorts.
    \item Categorize subjects into training, validation, or test cohorts to avoid slice level data leakage among cohorts. 
    \item Keep the ratio between good and poor quality images consistent in training and validation cohort.
\end{itemize}
\end{enumerate}

\subsubsection{Cross-Validation Scheme}
To evaluate the performance of MRIQC on unseen images for seen sites, MRIQC was trained with 5-fold cross validation schemes in both its inner and outer validation schemes~\cite{mriqc}. For evaluating performance of MRIQC on unseen sites, it was trained with nested cross-validation with a leave-one-site-out (LOSO) scheme since it demonstrated better generalization to new sites compared to k-fold cross-validation in~\cite{mriqc}. 

For evaluation of CNNQC performance on seen and unseen sites, CNNQC was trained with k-fold strategy with k=5.

\subsubsection{Performance Metrics}
The performance of predicting quality labels for brain MRI images was evaluated using accuracy, sensitivity~\cite{altman1994diagnostic} and balanced accuracy (BA)~\cite{brodersen2010balanced} - in line with methodologies adopted in prior studies~\cite{mriqc,JMRISheeba,3DCNNMRART}. We included sensitivity alongside accuracy to highlight any potential biases of a method towards specific classes given the strong class imbalance. The poor quality class in this work was the positive class, and hence sensitivity refers to predicting poor quality images as poor. 
\subsubsection{Comparison strategy}
The accuracy, sensitivity and BA values computed for MRIQC and CNNQC are compared in two following ways:
\begin{itemize}
    \item Relative comparison: The metric values for both methods compared to each other to demonstrate relative superiority of a method. 
    \item Absolute comparison: To identify how good a method performs. In absolute comparison a method was considered to be accurate if its accuracy value was greater than 0.55, was inaccurate if value if less than 0.45 and was uncertain otherwise. The method was sensitive if its sensitivity was greater than 0.5 and poor sensitivity otherwise. 
\end{itemize}

\section{Experimental Results}

\subsection{Evaluation on unseen sites}
\begin{figure}
    \centering
    \includegraphics[width=0.9\linewidth]{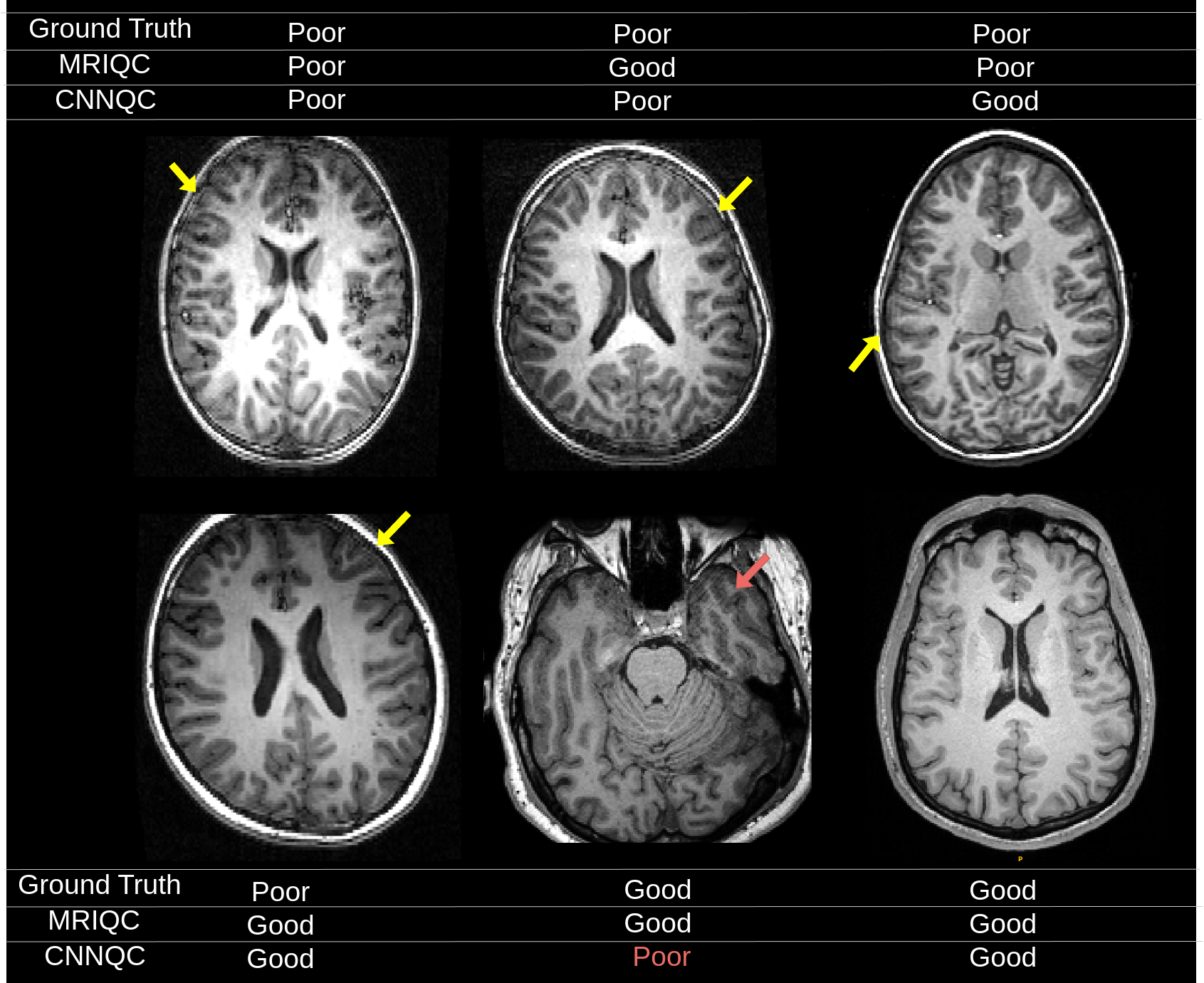}
    \caption{Example MRI images from various sites, their quality labels and the predicted quality labels by MRIQC and CNNQC. The yellow arrows indicate image artefacts, and red arrow indicate the potential imaging feature detected as artifact by CNNQC method.}
    \label{fig:mriimg}
\end{figure}

Example good and bad quality images are shown in Fig.~\ref{fig:mriimg}, demonstrating cases where the automated methods agree and disagree on. This also shows the variety in image contrast across sites.

\subsubsection{Accuracy and sensitivity}
\begin{figure}
    \centering
    \includegraphics[width=0.9\linewidth]{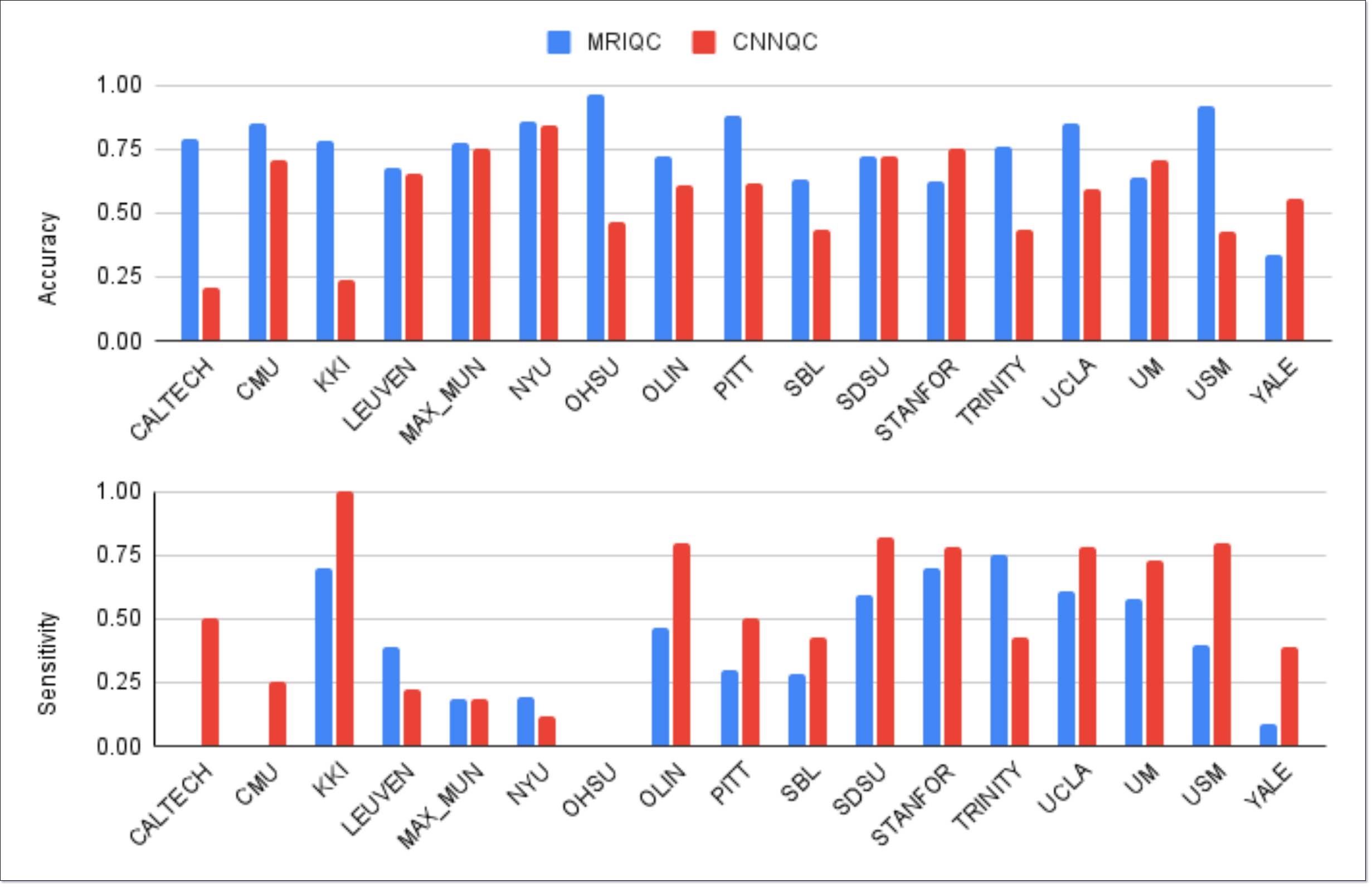}
    \caption{Bar plots of accuracy and sensitivity values.}
    \label{fig:rawAccSens}
\end{figure}

\begin{figure}
    \centering
    \includegraphics[width=0.8\linewidth]{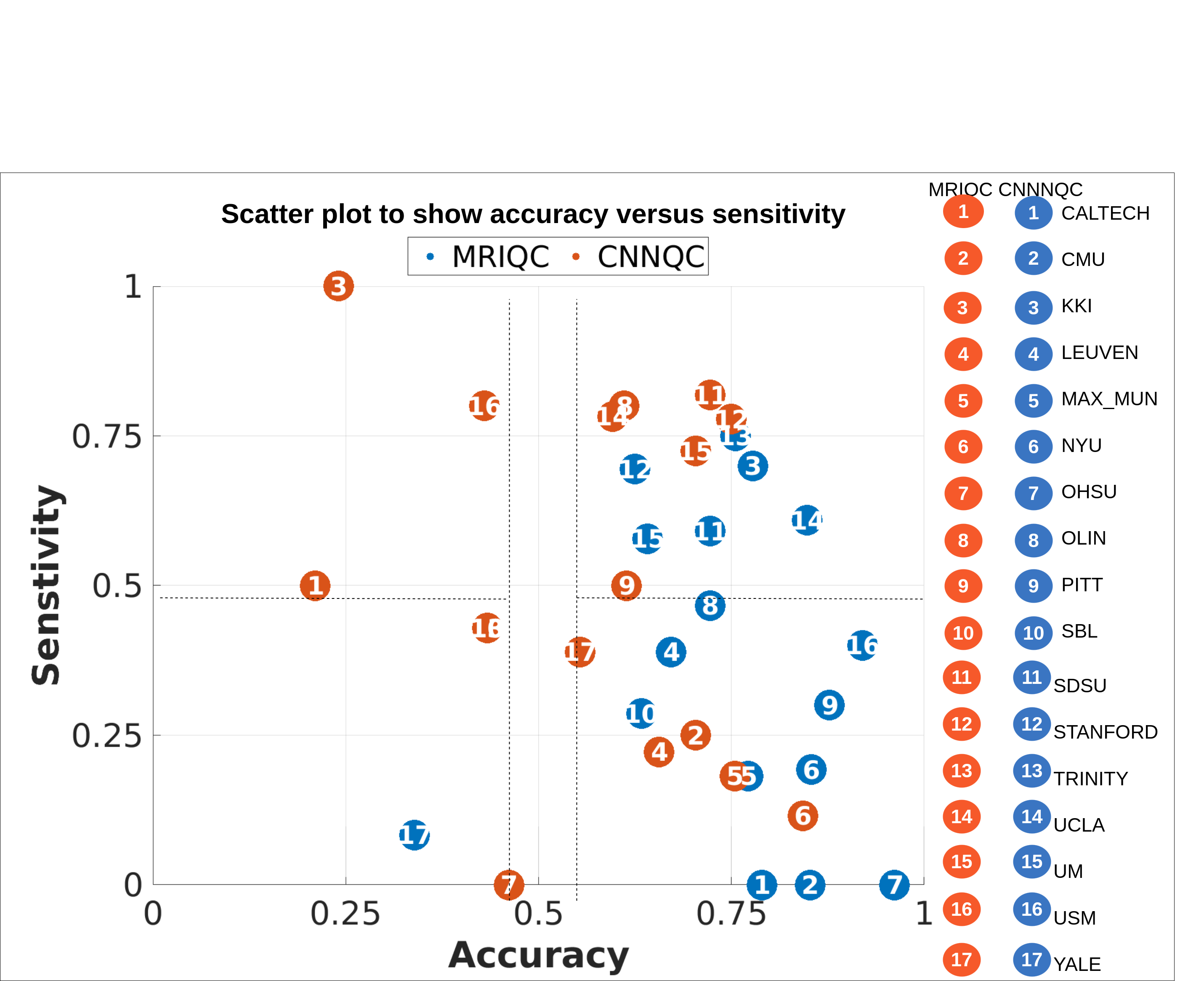} 
    \caption{Scatter plot between accuracy and sensitivity values.}
    \label{fig:relatingaccsens1}
\end{figure}

\begin{figure}
    \centering
    \includegraphics[width=\linewidth]{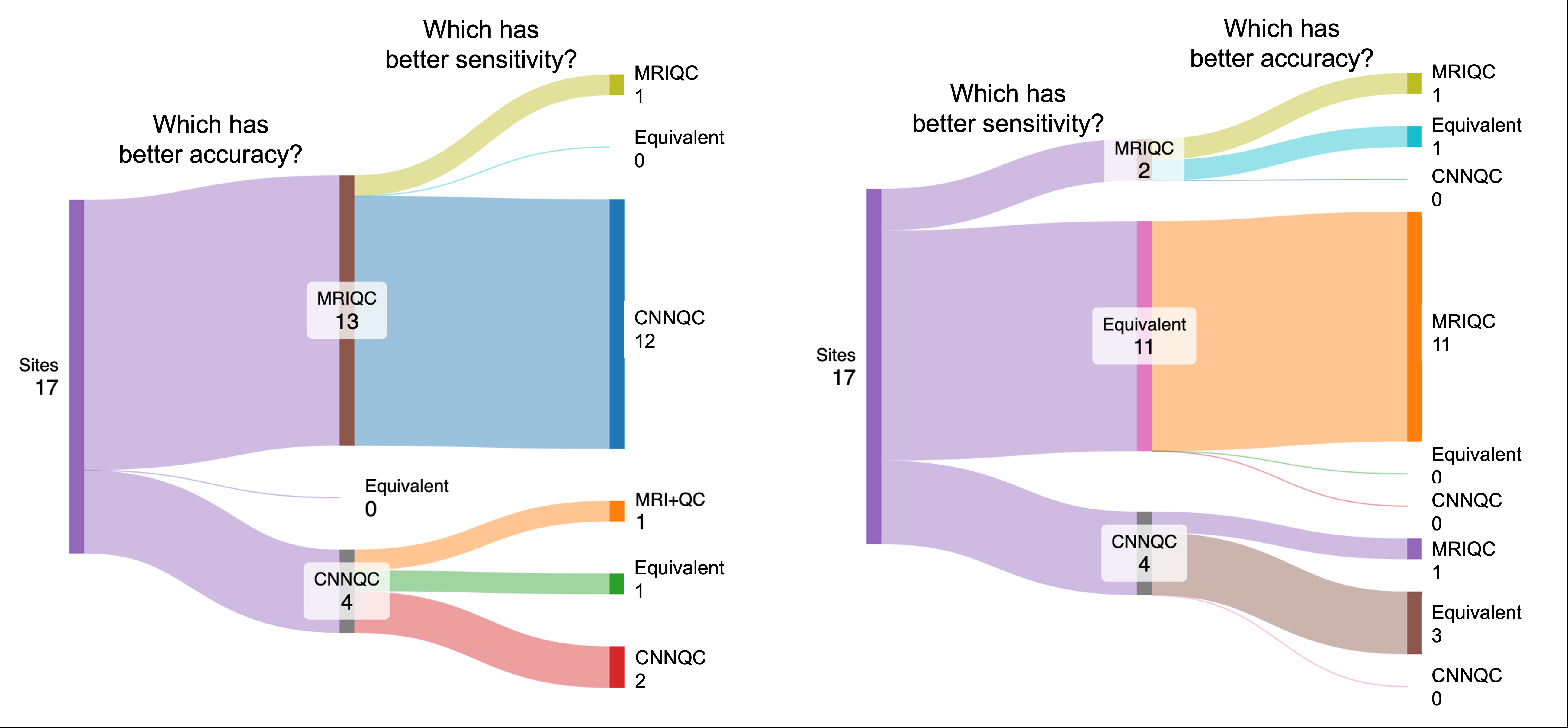} 
    \caption{Relative comparison of MRIQC and CNNQC in terms of accuracy and sensitivity}
    \label{fig:compareAcc}
\end{figure}

Fig.~\ref{fig:rawAccSens} shows the comparison between the two methods based on accuracy and sensitivity, and Fig.~\ref{fig:relatingaccsens1} the relation between accuracy and sensitivity. The values were compared with different comparison strategies in Fig.~\ref{fig:compareAcc}.

\begin{enumerate}[label=(\roman*)]
    
    \item Relative comparison: MRIQC has higher accuracy for 13 sites compared to CNNQC, while for 4 sites both methods have similar accuracy (difference < 0.1), Fig.~\ref{fig:compareAcc}. Among these 13 sites, CNNQC outperforms MRIQC in sensitivity for 12 of them. Out of the 4 sites with equivalent accuracy, CNNQC has higher sensitivity for 2 sites than MRIQC. This suggests that, although MRIQC is more accurate, CNNQC tends to have relatively higher sensitivity.
    \item Absolute comparison: Results for accuracy and sensitivity are listed in Table~\ref{tab:absaccsens}. MRIQC has accuracy>0.55 for 16 sites whereas CNNQC achieves high accuracy for 11 sites. MRIQC has sensitivity >0.5 for 6 sites and CNNQC provides high sensitivity for 7 sites.
    \item Relation between accuracy and sensitivity:  MRIQC displays a bias towards classifying images as good quality, indicated by higher accuracy across most sites but lower sensitivity. In an absolute comparison, MRIQC's performance aligns more with a unimodal distribution, suggesting a consistent preference for one class. In contrast, CNNQC has relatively bimodal distribution meaning it performs better for both output classes - good or bad.
\end{enumerate}
Combining the relative and absolute comparison shows that MRIQC might seem more accurate and CNNQC seems more sensitive to bad quality scans. Analyzing the absolute accuracy and sensitivity values suggests both methods have poor overall performance and thus show low generalisation.

\begin{table}[]
\caption{Absolute comparison on the basis of accuracy and sensitivity}
\begin{tabular}{|c|c|c|c|}
\hline
\textbf{Accuracy} & \textbf{Sensitivity} & \textbf{MRIQC} & \textbf{CNNQC} \\ \hline
Low               & Low                   & 1              & 3              \\ \hline
Low               & High                  & 0              & 2              \\ \hline
High              & Low                   & 10             & 6              \\ \hline
High              & High                  & 6              & 5              \\ \hline
Uncertain         & -                     & 0              & 1              \\ \hline
\end{tabular}
\label{tab:absaccsens}
\end{table}

%
%

\subsubsection{Balanced Accuracy}
\begin{figure}
    \centering
    \includegraphics[width=0.9\linewidth]{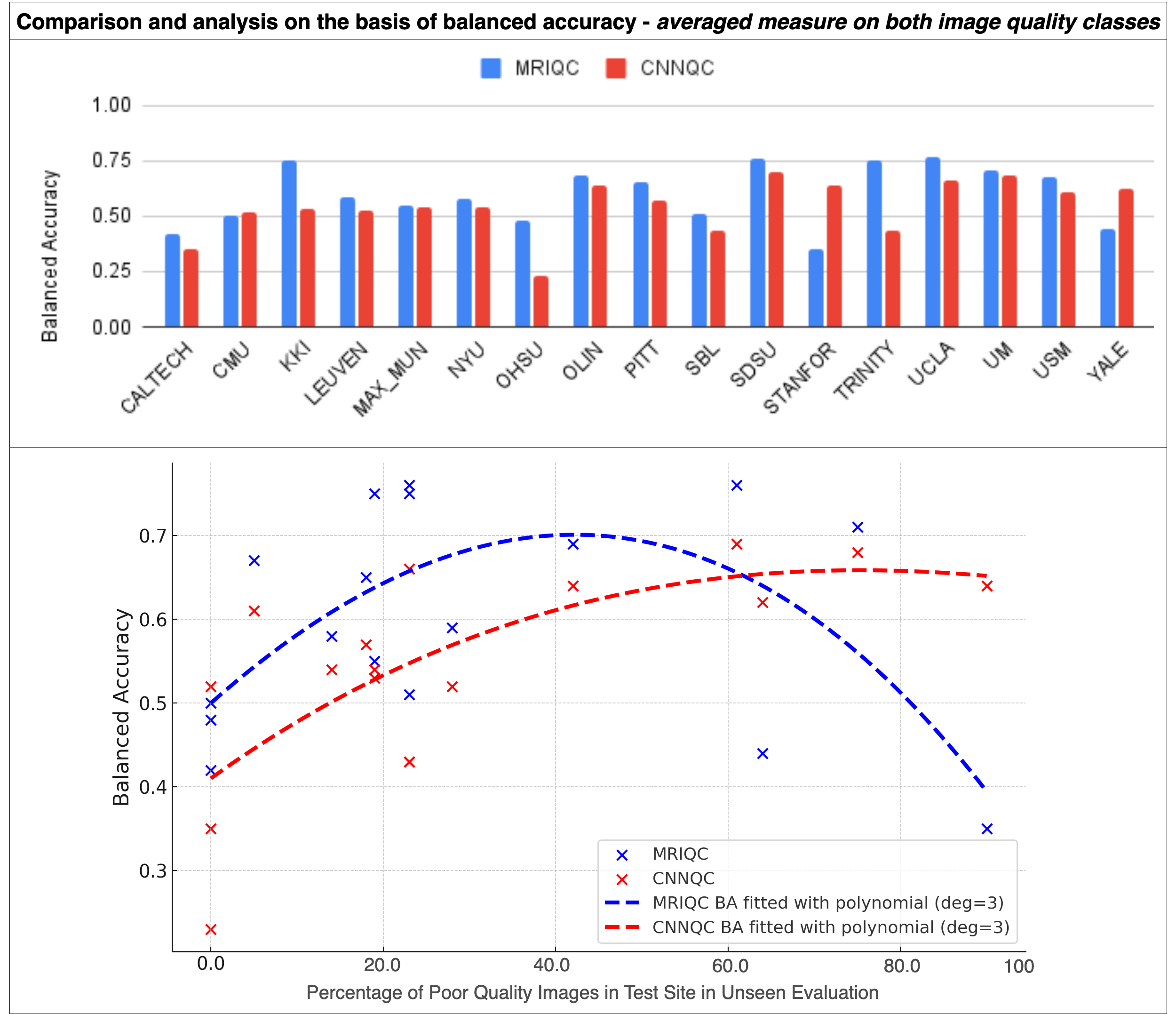}
    \caption{Above: Comparison of balanced accuracy values for MRIQC and CNNQC. Below: Balanced accuracy of MRIQC and CNNQC plotted against the percentage of poor quality images per test site. Polynomial fits of degree three are overlaid to illustrate trends.}
    \label{fig:ba}
\end{figure}

Graphical representation of BA values for both methods and their comparison is shown in Fig.~\ref{fig:ba}. In a relative comparison, 11 sites show equivalent BA (difference <0.1), with MRIQC having a higher BA for four sites, and CNNQC for two sites. Second row in Fig.~\ref{fig:ba} shows the relation between BA and percentage of poor quality scans. MRIQC demonstrates superior performance when the dataset contains a higher proportion of good quality images, whereas CNNQC tends to perform better when poor quality scans dominate. Both methods perform similarly when the proportion of good quality images lies between 40\% and 60\%. Importantly, there is strong variability in MRIQC's performance with higher percentage poor quality scans, while CNNQC is more stable at this range. In the absolute comparison, both methods provide BA>0.5 for 12 sites but the highest BA does not exceed 0.76 indicating need for improvement. 

Both MRIQC and CNNQC methods achieve their highest Balanced Accuracy (BA) for the SDSU site, i.e., 0.76 and 0.69, respectively. Interestingly, the specificity and sensitivity scores for SDSU site are reversed for both methods while having equal accuracy of $\sim$0.72. MRIQC shows a specificity of 0.92 and sensitivity of 0.59, while CNNQC presents the reverse, with an specificity of 0.57 and sensitivity of 0.81. This contrast highlights the distinct characteristics and potential performance tendencies of each method.

\subsection{Evaluation on seen sites}
The performance metrics for both methods using both data labeling sets are mentioned in Table~\ref{tab:seen}. CNNQC evaluated with ABIDE ratings present similar values as reported in~\cite{JMRISheeba} demonstrating successful validation and replication of the method. 

\begin{table}[t!]
    \centering
    \begin{tabular}{|c|c|c|c|c|}
    \hline
      Approach & Ratings & Accuracy & Sensitivity & BA\\ \hline
         CNNQC & MRIQC &  0.70&  0.91& 
         0.76\\ \hline  
         MRIQC & MRIQC &  0.85 & 0.75 & 0.82 
         \\ \hline
     CNNQC & ABIDE & 0.84 & 0.87 & 0.86\\ \hline
   MRIQC & ABIDE & 0.94 & 0.52 & 0.75\\ \hline

    \end{tabular}
    \caption{Comparison of approaches' performance evaluated on new images from seen sites.}
    \label{tab:seen}
\end{table}

\subsubsection{Accuracy and Sensitivity:}
In the absolute comparison, CNNQC is accurate and sensitive whereas MRIQC is accurate but with lower sensitivity. This observation is in concordance with the poor recall with good precision in the MRIQC paper for the held-out dataset~\cite{mriqc}.

\subsubsection{Balanced Accuracy:}
CNNQC provides relatively better BA values than MRIQC.

\section{Summary}

This study investigates whether DL methods surpass traditional ML techniques that utilize handcrafted features in AQC of brain MRI images. Our findings reveal that both approaches are essentially equivalent in performance and notably underperform when applied to test images from new scanners or sites. This underscores the critical need for future AQC methods to focus on achieving greater generalizability. Given equivalent performance between two methods, the computational run-time advantage of deep learning based methods make that a preferred method for efficient deployment and widespread use. Irrespective of choice of method, one has to be cautious of the challenges in each method. For CNNQC these are: 1) Pre-processing: Even though deep learning requires less pre-processing, there are constraints on spatial resolution based on the trained model, with data cropped or padded as per the input shape of trained CNN; 2) Overfitting: Even with fewer parameters than 3D-CNN, 2D-CNN tends to overfit the dataset when evaluated using class-weighted loss because of noise in labels, misaligned image volumes, intensity histogram changes. MRIQC method uses Area under curve (AUC) as optimisation metric to address the class imbalance problem during training, but our results suggest residual inclination toward the majority class. Although we used a multi-site, multi-vendor public dataset, contrast variability in routine clinical MRI is likely greater; therefore, further method development and rigorous external validation—ideally prospective and multi-centre—are needed to ensure the widespread application of the developed methods. 


\section*{Acknowledgements}
We are grateful to the Rosetrees Trust (UCL-IHE-2020\textbackslash101) and the National Institute for Health Research University College London Hospitals Biomedical Research Centre for supporting this work. The authors acknowledge the facilities and scientific and technical assistance of the National Imaging Facility, a National Collaborative Research Infrastructure Strategy (NCRIS) capability, at the Centre for Microscopy, Characterisation, and Analysis, the University of Western Australia.

\bibliographystyle{ACM-Reference-Format}
\bibliography{sample-base}

\newpage

\end{document}